# Rumor Detection and Classification for Twitter Data

Sardar Hamidian and Mona Diab
Department of Computer Science
The George Washington University
Washington DC, USA
Email: sardar@gwu.edu, mtdiab@gwu.edu

*Abstract*—With the pervasiveness of online media data as a source of information, verifying the validity of this information is becoming even more important yet quite challenging. Rumors spread a large quantity of misinformation on microblogs. In this study we address two common issues within the context of microblog social media. First, we detect rumors as a type of misinformation propagation, and next, we go beyond detection to perform the task of rumor classification (RDC). We explore the problem using a standard data set. We devise novel features and study their impact on the task. We experiment with various levels of preprocessing as a precursor to the classification as well as grouping of features. We achieve an F-Measure of over 0.82 in the RDC task in a mixed rumors data set and 84% in a single rumor data set using a two step classification approach.

## I. INTRODUCTION

Social media is currently a place where massive data is generated continuously. Nowadays, novel breaking news appear first on microblogs, before making it through to traditional media outlets. Hence, microblogging websites are rich sources of information which have been successfully leveraged for the analysis of sociopragmatic phenomena, such as belief, opinion, and sentiment in online communication. Twitter [27] is one of the most popular microblogging platforms. It serves as one of the foremost goto media for research in natural language processing (NLP), where practitioners rely on deriving various sets of features leveraging content, network structure, and memes of users within these networks. However, the unprecedented existence of such massive data acts as a double edged sword, one can easily get unreliable information from such sources, and it is a challenge to control the spread of false information either maliciously or even inadvertently. The information seeker is inundated with an influx of data. Most importantly, it is hard to distinguish reliable information from false information, especially if the data appears to be formatted and well structured [9] [24]. The problem is exacerbated by the fact that many information seekers believe that anything online in digital form is true and that the information is accurate and trustworthy; although, it is well known that a lot of the information on the web could be false or untrue. This is especially crucial in cases of emergencies. For example, by simply hitting the Re-tweet button on Twitter, within a fraction of a second, a piece of information becomes viral almost instantly. There are widely varying definitions of the term "rumor". We adopt the following definition of rumor: a rumor could be both true or false. A rumor is a claim whose truthfulness is in doubt and has no clear source, even if its ideological or partisan origins and intents are clear [2].

In verifying the accuracy of claims or events online, there are four major aspects that could be checked: *Provenance*, the original piece of content; *Source*, who uploaded the content; *Date-and-location*, when and where the content was created [22]. Analyzing each of these items individually plays a key role in verifying the trustworthiness of the data.

In this paper, we address the problem of detecting rumors in Twitter data. We start with the motivation behind this research, and then the history of different studies about rumors is overviewed in Section 2. Next, in Section 3, the overall pipeline is exposed, in which we adopt a supervised machine learning framework with several feature sets, and finally in Section 4, we compare our results to the current state of the art performance on the task. We show that our approach yields comparable and even superior results to the work to date.

## II. RELATED WORK

Psychologists studied the phenomenon of rumors from various angles. First studies were carried out in 1902 by German psychologist and philosopher, William Stern, and later in 1947 by his student Gordon Allport, who studied how stories get affected in their lifecycle [10]. In 1994, Robert Knapp published "A Psychol-ogy of Rumors", which comprised of a collection of more than a thousand rumors propagated during World War II. In his work, the rumor is what was transmitted by word of mouth and bore information about a person, an event, or a condition, which fulfilled the emotional desires of the public [11]. In 1948, Allport and Postman [12] studied the behavior of rumors and how one rumor reflects leveling, sharpening, and assimilation behavior in its propagation. Related studies in political communication conducted by Harsin [2] presented the idea of the "Rumor Bomb". For Harsin, a "Rumor Bomb" spreads the notion of the rumor into a political communication concept. In other research, Kumar and Geethakumari [5] explore the use of theories in cognitive psychology to estimate the spread of disinformation, misinformation, and propaganda across social networks. There are several studies about the behavior of misinformation and how they are distinguished in a microblog network. For example, Budak took a step further [4] investigating how to overcome the spread of misinformation by applying an optimized limitation campaign to counteract the effect of misinformation.

From an NLP perspective, researchers have studied numerous aspects of credibility of online information. For example, Ghaoui [3] detects rumors within specialized domains, such as trustworthiness, credibility assessment and information verification in online communication. Modeling and monitoring the social network as a connected graph is another approach. Seo [6] identifies rumors and their corresponding sources by observing which of the monitoring nodes receive the given information and which do not. Another relevant work, Castillo [23], applied the time-sensitive supervised approach by relying on the tweet content to address the credibility of a tweet in different situations. The most relevant related work to ours is that reported in [1], which addresses rumor detection in Twitter using content-based as well as microblog-specific meme features. However, differences in data set size and number of classes (rumor types) render their results not comparable to ours. Moreover, Qazvinian et al. [1] suggest label-dependent features in creating their User-based (USR) and URL features, which is only possible by having the input data labeled for being a rumor or not. In other words, labeled data is used for creating the language model (LM) with USR and URL features, and the trained LM is then used for extracting the value of each feature. In our study, we propose a totally label-independent method for feature generation that relies on the tweet content, and boosts

TABLE I. LIST OF ANNOTATED RUMORS [1]

| Rumor | Rumor Reference | # of tweets |
|---|---|---|
| Obama | Is Barack Obama muslim? | 4975 |
| Michele | Michelle Obama hired many staff members? | 299 |
| Cellphone | Cell phone numbers going public? | 215 |
| Palin | Sarah Palin getting divorced? | 4423 |
| AirFrance | Air France mid-air crash photos? | 505 |

our model in a realtime environment.

## III. APPROACH

We addressed the problem of rumor detection and classification (RDC) within the context of microblog social media. We focused our research on Twitter data due to the availability of annotated data in this genre, in addition to the above mentioned interesting characteristics of microblogging, and their specific relevance to rumor proliferation.

### A. Data

Qazvinian et al. [1] published an annotated Twitter data set for five different 'established' rumors as listed in Table I. The general annotation guidelines are presented in Table II.

TABLE II. RUMOR DETECTION ANNOTATION GUIDELINES

| 0 | If the tweet is not about the rumor |
|---|---|
| 11 | If the tweet endorses the rumor |
| 12 | If the tweet denies the rumor |
| 13 | If the tweet questions the rumor |
| 14 | If the tweet is neutral |
| 2 | If the annotator is undetermined |

The following examples illustrate each of the annotation labels from the Obama rumor collection.

**0**: 2010-09-24 15:12:32 , nina1236 , Obama: Muslims 2019 Right To Build A Manhattan Mosque: While celebrating Ramadan with Muslims at the White House, Presi... http://bit.ly/c0J2aI

**11**: 2010-09-28 18:36:47 , Phanti , RT @IPlantSeeds: Obama Admits He Is A Muslim http://post.ly/10Sf7 - I thought he did that before he was elected.

**12**: 2010-10-01 05:00:28 , secksaddict , barack obama was raised a christian he attended a church with jeremiah wright yet people still beleive hes a muslim

**13**: 2010-10-09 06:54:18 , affiliateforce1 , Obama, Muslim Or Christian? (Part 3) http://goo.gl/fb/GJtsJ

**14**: 2010-09-28 22:22:40 , OTOOLEFAN , @JoeNBC The more Obama says he's a Christian, the more right wingers will say he's a Muslim."

**2**: 2010-10-05 17:37:04 , zolqarnain , Peaceful Islam- Muslims Burn CHURCH in Serbia: http://wp.me/p121oH-1ir OBAMA SILENT #politics #AACONS #acon #alvedaking #women #news #tcot

Table III shows statistics for the annotated tweets corresponding to each of the five rumors. The original data set as obtained from [1] did not contain the actual tweets for both Obama and Cellphone rumors, but they only contained the tweet IDs. Hence, we used the Twitter API for downloading the specific tweets using the tweet ID. Accordingly, the size of our data set is different from that of [1] amounting to 9000 tweet in total for our experimentation.

TABLE III. LIST OF ANNOTATED TWEETS PER LABEL PER RUMOR

| Rumor | 0 | 11 | 12 | 13 | 14 | 2 | Total |
|---|---|---|---|---|---|---|---|
| Obama | 945 | 689 | 410 | 160 | 224 | 1232 | 3666 |
| Michelle | 83 | 191 | 24 | 1 | 0 | 0 | 299 |
| Palin | 86 | 1709 | 1895 | 639 | 94 | 0 | 4423 |
| Cellphone | 92 | 65 | 3 | 3 | 3 | 0 | 166 |
| Air France | 306 | 71 | 114 | 14 | 0 | 0 | 505 |
| Mix | 1512 | 2725 | 2452 | 817 | 321 | 1232 | 9059 |

### B. Experimental Conditions

We approached RDC in a supervised manner and investigated the effectiveness of multi step classification with various sets of features and preprocessing tasks versus a single step detection and classification approach. In the single-step classification for RDC, we performed detection and classification simultaneously as a 6-way classification task among the six classes in the labeled data, as shown in Table II, by retrieving the tweets as Not Rumor(0), Endorses Rumor(11), Denies Rumor(12), Questions Rumor(13), Neutral(14), and Undetermined tweets(2). In the two-step classification set up, an initial 3-way classification task is performed among the following groups of fine grained labels (0, Not Rumor), (2, Undetermined tweet), and the compound (11-14, Rumor) labels. This is followed by a 4-way classification step for the singleton labels, (11, Endorsing the Rumor), (12 Denys the Rumor), (13, Questions the Rumor), and (14, Neutral about the Rumor). In the second step, we took out class 0 and 2 tweets from the training data set and only classified the tweets from the test data set, which had been classified as rumor in the first step. The underlying motivation of our effort in designing the single-step and two-step classification is to investigate the performance of each technique in order to solve two problems. First, classifying tweets as 'Rumor' and 'Not Rumor', which can assist users to distinguish the type of tweets. Second, classifying the rumor type that the tweet endorses, denies, questions or is neutral. Although in both problems we investigated the rumor, these two problems are different. Our two-step model pipeline is dynamic in a way that the output of the the first step (Rumor Detection) is the input data set for the next step (Rumor Type Classification). We also designed a new set of pragmatic features along with updating the set of features in Twitter and network-specific category, which could boost the overall performance in our pipeline.

### C. Machine Learning Frameworks

For our experiment we applied J48, a discriminative classifier that utilizes decision trees and supports various types of attributes. WEKA platform [25] is used for training and testing the proposed models in our pipeline.

TABLE IV. FINAL LIST OF USED FEATURES. '*' MARKED FEATURES ARE THE APPENDED SET OF FEATURES

| | ID | Value |
|---|---|---|
| Twitter and Network Specific | * Time | Binary |
| | * Hashtag | Binary |
| | Hashtag Content | String |
| | URL | Binary |
| | Re-tweet | Binary |
| | *Reply | Binary |
| | User ID | Binary |
| Content | Content Unigram | String |
| | Content Bigram | String |
| | Pos Unigram | String |
| | Pos Bigram | String |
| Pragmatic | *NER | String |
| | *Event | String |
| | *Sentiment | String |
| | *Emoticon | Binary |

### D. Feature Sets

We experimented with content, network, and social meme features. We extended the number of features by including the pragmatic attributes. We employed all the features proposed in [1] in addition to developing more pragmatic attributes as well as additional network features. For network and meme features, we explicitly modeled source and timestamped information and for pragmatic features we proposed NER, Event, Sentiment, and Emoticon. Table IV lists all the features for the RDC task and marked the new features with "*".

*1) Content Features:* This set of features is developed using tweet content. We applied various preprocessing granularity levels to measure the impact of preprocessing on the RDC task.

*a) Unigram-Bigram Bag of Words (BOW):* Similar to the content lexical features proposed in [1], we used a bag of words feature set comprising word unigrams and bigrams. We employed the WEKA's String To Word Vector along with N-gram tokenizer for creating this feature set with the TF-IDF weighting factor as the matrix cell content corresponding to each feature. We also generated the lemma form of the words in the tweets using WordNet [19] lemmatization capability. Accordingly, we created four feature sets: unigram tokenized word form, unigram lemma form, bigram tokenized word form, and bigram lemma form.

*b) Part of Speech (POS):* POS tagging for social media is challenging since the text genre is informal and quite noisy. We relied on the CMU Twitter POS tagger [7]. The feature values are set to a binary 0 or 1, corresponding to unseen or observed.

*2) Pragmatic Features:* In an extension to the features proposed by [1], we further explored the explicit modeling of pragmatic features to detect favorable and unfavorable opinions toward specific subjects (such as people, organizations). Applying this set of features offers enormous opportunities for detecting the type of rumors [21].

*a) Sentiment:* There are a wide variety of features for sentiment classification on Twitter data sets that have been studied in various publications. We believe that polarity of a tweet could be an informative factor to extract user's opinion about each rumor. For tagging the sentiment polarity of a tweet we applied the Stanford Sentiment system [18]. We preprocessed the data by removing punctuations, URL, "RT", and lowercased the content. Each tweet is tagged with one of the following sentiment labels; ***Very Positive, Positive, Neutral, Negative, or Very Negative***.

*b) Emoticon:* Another pragmatic cue is Emoticon. Studies on modeling and analyzing microblogs, which explicitly use emoticon as a feature, show its impact on classification [17]. We used the list of popular emotions described in Wikipedia [26]. We manually designated and labeled the list of entries as either expressing Positive (2), Negative (1), or Neutral (0) emotions.

*c) Named-Entity Recognition (NER):* We employed Twitter NLP tools [20] to explicitly extract information about named-entities, such as Location, Person, Organization, etc. In this paper we show how modeling NER has an explicitly positive impact on performance.

*d) Event:* Extracting the entity *Obama* and the event phrase *praises in connection with Muslims* is much more informative than simply extracting *Obama*. We utilized the same Twitter NLP tools [20] for tagging event labels.

*3) Network and Twitter Specific Features:* Relying on Twitter specific memes, we expanded features listed in [1] by adding time and network behavior features, such as Reply.

*a) Time:* It is quite remarkable that social networks spread news so fast. In a similar task to [13] we analyzed the process of rumor expansion on Twitter in our data set. Both the structure of social networks and the process that distributes the news lead to a piece of news becoming viral instantaneously. We labeled and ranked all the days based on the number of tweets posted in a day. We modeled the tweet creation-time attribute. We also observed that more than 90% of rumors are posted during the five most busiest days in the collected data set. Figure 1 shows the results of tabulating time frequency of the rumors in the Palin rumor data set and how the number of rumors changed within a six month period. Accordingly, we designated two labels for the time feature: Busy Day or Regular Day, depending on what type of day tweets were (re)tweeted.

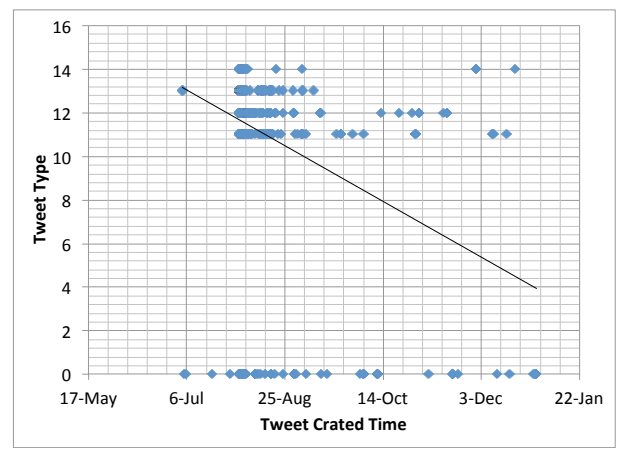

Figure 1. tweet Distribution for the Palin Rumor collection within a 6 month period

*b) Reply, Re-tweet, User ID:* Replying and retweeting in microblogs are revealing factors in judging user's trustworthiness when it comes to relaying information [14]. For example, *User A* is more likely to post a rumor than *User B* if *User A* has a history of retweeting or replying to *User C* who also has a rumor spreading history. Investigating the credibility of users is an expensive and almost impossible task, but it is doable when we only want to investigate a specific story. For example, knowing the limited number of users who have a history of posting rumors could be a hint to detect the large number of users that follow, retweet or reply to those tweets.

TABLE V. NUMBER OF FEATURES AND LABELS USED IN SINGLE STEP AND TWO STEP CLASSIFICATIONS

| | 1st Step | 2nd Step |
|---|---|---|
| Method | Labels | Labels |
| (SRDC) 6-way classification | (0)(11)(12)(13)(14)(2) | |
| (TRDC) 3-way (1st step) — 4-way (2nd step) classification | (0)(2)(11-14) | (11)(12)(13)(14) |

*c) Hashtag:* Hashtags serve as brief explanations of the tweet content [16]. We extracted hashtags in the labeled data set. Tweets with no hashtags are assigned a value 0 to their hashtag feature dimension, and tweets containing hashtag(s), received a 1 in the hashtag dimension. Additionally, we added all the observed hashtags as feature dimensions, thereby effectively identifying the tweets that share the same hashtag. For compound hashtags, we used a simple heuristic. If the hashtag contained an uppercase character in the middle of the hashtag word, then we split it before the uppercase letter. For instance, **#SarahDivorce** is separated into two hashtags and converted to **Sarah** and **Divorce**. We then modeled both compound and separated hashtags as hashtag feature dimensions.

*d) URL:* Twitter users share URLs in their tweets to refer to external sources as an authentic proof (a source of grounding) to what they share. All URLs posted in tweets are shortened into 22 characters using the Twitter t.co service. Analyzing the URL is an expensive task and requires a huge source of information to verify the content of the shared URL. We excluded all URLs but we modeled their presence as a binary feature.

## IV. EXPERIMENTAL DESIGN

All the experiments are designed, performed, and evaluated based on various experimental settings and conditions, all elaborated in this section.

### A. Data

We experimented with three data sets: the two largest rumor sets, Obama and Palin, and a mixed data set (MIX) which comprises all the data from the five rumors. We splited each of the three data sets into 80% train, 10% development, and 10% test.

### B. Experimentation Platform

All experimentations were carried out using the WEKA-3-6 platform [25].

### C. Baselines

We adopted two baselines: Majority and limiting the features to the set of features proposed in [1], which are Content, Hashtag-Content, URL, Re-tweet, and User ID. As the name indicates, the Majority baseline assigns the majority label from the training data set to all the test data.

### D. Experimental Conditions and Evaluation Metrics

We had two main experimental conditions: single-step RDC (SRDC) and a two-step RDC (TRDC). We employed the set of 15 features listed in Table IV. Information about SRDC and TRDC is illustrated in Table V. In the development phase multiple settings and configurations were performed on the development data set for tuning, then the models that achieved the highest performance were used on the test set. Evaluating the performance of the proposed technique in rumor detection should rely upon both the number of relevant rumors that are selected (recall) and the number of selected rumors that are relevant (precision). Hence, we calculated F-measure, a harmonic mean of precision and recall due to its bias as an evaluation metric. TableVI shows the F-measure value for the different settings on the test set.

## V. RESULTS AND ANALYSIS

In this section the impacts of different experimental conditions are investigated.

### A. SRDC and TRDC

By studying the results in Table VI, it can be observed that TRDC significantly outperforms SRDC, since TRDC achieves an F measure of 82.9% compared to 74% in SRDC for the MIX data set, and 85.4% for the Obama data set compared to a 71.7% in SRDC. By comparing the F-Measure with the Majority baseline and the features proposed in [1](VAR11) as the second baseline, we could explicitly see how applying the proposed methodology and set of features enhances the overall performance in certain rumors, and also leads to acceptable performance in the MIX data set.

TABLE VI. F-MEASURE RESULTS OF SRDC AND TRDC METHODS EMPLOYING 15 FEATURES AND VAR11 FEATURES

| data set | Method | Our 15 Feat. | VAR11 Feat. |
|---|---|---|---|
| MIX | Majority | 0.30 | |
| | SRDC | **0.743** | **0.748** |
| | TRDC | **0.83** | **0.83** |
| 1-3 Obama | Majority | 0.33 | |
| | SRDC | **0.717** | 0.705 |
| | TRDC | **0.854** | 0.844 |
| Palin | Majority | 0.46 | |
| | SRDC | **0.754** | 0.748 |
| | TRDC | **0.79** | 0.70 |

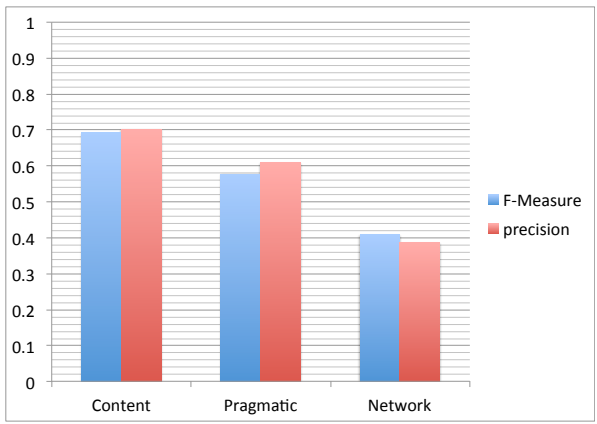

Figure 2. The Average F-measure and precision of SRDC and TRDC classifications employing each group of features: Content, Pragmatic, Network

### B. Impact of Feature Set

In this experiment, we assessed the performance of different groups of features individually. Figure 2 shows the average F-measure and precision of SRDC and TRDC by employing the Content, Pragmatic and Network sets of features. As shown in Figure 2, employing the Content set of features yields the overall best precision. In contrast to the other features, the network feature set had the minimum impact on our classification.

### C. Impact of Preprocessing

As mentioned above, we applied various levels of preprocessing to the content of tweets such as stemming, lemmatization, punctuation removal, lowercasing, and stop words removal. We measured the impact of applying such preprocessing versus no preprocessing. Figure 3 illustrates that accuracy doesn't benefit from preprocessing and results in the loss of valuable information.

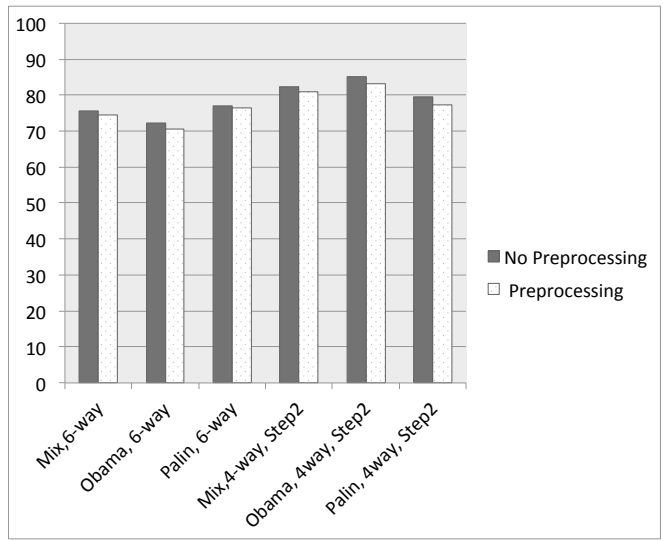

Figure 3. The overall accuracy in different experiments with and without preprocessing

## VI. CONCLUSIONS AND FUTURE WORK

In this paper, we study the impact of a single-step (SRDC) 6 way classification versus a two-step classification (TRDC). Our contributions in this paper are two-fold: (1) We boosted the pipeline by decoupling the rumor detection from the classification task. We proposed an automated TRDC pipeline that employs the results from the rumor detection step and performs the classification task upon data and leads to promising results in comparison to SRDC. (2) We employed a new set of meta linguistic and pragmatic features, which leads and performs the experiments with and without preprocessing on the textual content. We achieved the F-Measure of more than 0.82 and 0.85 on a mixed and the Obama rumor data sets, respectively. Our proposed features achieved better performance compared to the state of the art features proposed in [1]. Our study however suggests that our pipeline does not benefit from preprocessing which might be attributed to the weakness of the tools used for processing twitter content at this stage. We are planning to expand the proposed methodology to streaming tweets. Having a limited amount of labeled data, we are investigating means of augmenting the training data with noisy data in a semisupervised framework.

ACKNOWLEDGEMENT

This paper is based upon work supported by the DARPA DEFT Program.